\documentclass[aps,pre,twocolumn,showpacs,superscriptaddress]{revtex4}

\pdfoutput=1

\usepackage{amsmath,amssymb}
\usepackage{bm}
\usepackage{graphicx}

\usepackage{color}

\usepackage{ulem}

\newcommand{\rev}[1]{\textcolor{black}{#1}}

\begin{document}

\title{
Tunable dynamic response of magnetic gels: impact of structural properties and magnetic fields
}

\author{Mitsusuke Tarama}
\affiliation{Department of Physics, Kyoto University, Kyoto 606-8502, Japan
}
\affiliation{Institut f\"ur Theoretische Physik II: Weiche Materie, Heinrich-Heine-Universit\"at D\"usseldorf, D-40225 D\"usseldorf, Germany
}
\affiliation{Institute for Solid State Physics, University of Tokyo, Kashiwa, Chiba 277-8581, Japan
}

\author{Peet Cremer}
\affiliation{Institut f\"ur Theoretische Physik II: Weiche Materie, Heinrich-Heine-Universit\"at D\"usseldorf, D-40225 D\"usseldorf, Germany
}

\author{Dmitry Y.\ Borin}
\affiliation{Technische Universit\"at Dresden, Institute of Fluid Mechanics, D-01062, Dresden, Germany
}

\author{Stefan Odenbach}
\affiliation{Technische Universit\"at Dresden, Institute of Fluid Mechanics, D-01062, Dresden, Germany
}

\author{Hartmut L\"owen}
\affiliation{Institut f\"ur Theoretische Physik II: Weiche Materie, Heinrich-Heine-Universit\"at D\"usseldorf, D-40225 D\"usseldorf, Germany
}

\author{Andreas M.\ Menzel}
\email{menzel@thphy.uni-duesseldorf.de}
\affiliation{Institut f\"ur Theoretische Physik II: Weiche Materie, Heinrich-Heine-Universit\"at D\"usseldorf, D-40225 D\"usseldorf, Germany
}

\date{\today}

\begin{abstract}
Ferrogels and magnetic elastomers feature mechanical properties that can be reversibly tuned from outside through magnetic fields. Here we concentrate on the question how their dynamic response can be adjusted. The influence of three factors on the dynamic behavior is demonstrated using appropriate minimal models: first, the orientational memory imprinted into one class of the materials during their synthesis; second, the structural arrangement of the magnetic particles in the materials; and third, the strength of an external magnetic field. To illustrate the latter point, structural data are extracted from a real experimental sample and analyzed. 
Understanding how internal structural properties and external influences impact the dominant dynamical properties helps to design materials that optimize the requested behavior. 
\end{abstract}



\maketitle


\section{Introduction} \label{sec:Introduction}


Often the internal dissipation in soft matter systems is sufficiently large so that their dynamics can be considered as overdamped. For instance the motion of dispersed colloidal particles is dominated by the friction with the surrounding liquid \cite{ivlev2012complex}. Another example is the dynamics of polymer chains in melt or solution, described in a first approach by the famous Rouse and Zimm models \cite{rouse1953theory,zimm1956dynamics}. 
Apart from that, in polymeric systems the dynamic behavior is often dominated by relaxation processes. The reason is found in the large size of their building blocks. A long time is necessary for conformational rearrangements to adjust to changes in their environment \cite{strobl1997physics}. Frequently, the slower processes are the ones that strongly influence the macroscopic behavior. 

Here, we consider the combination of the two materials mentioned above in the form of ferrogels or magnetic elastomers \cite{filipcsei2007magnetic}. In this case, magnetic colloidal particles are embedded into a crosslinked polymeric matrix. Qualitatively different kinds of this ``embedding'' can be achieved by different protocols of synthesis. On the one hand, the magnetic particles can simply be enclosed in mesh pockets of the polymer network \cite{filipcsei2007magnetic}. This allows a certain degree of freedom for particle reorientations. 
On the other hand, via surface functionalization, the magnetic particles can serve as crosslinkers and thus become part of the polymer mesh \cite{frickel2011magneto,messing2011cobalt,weeber2012deformation}. Then, restoring torques hinder reorientations of the particles. We use the term ``orientational memory'' to refer to this situation \cite{annunziata2013hardening}. 

From the internal architecture of these materials it is obvious that their magnetic and mechanical properties are strongly coupled to each other. This is what makes them interesting from both an academic and an application point of view. For example, the mechanical properties, such as the mechanical elastic modulus, can be tuned and adjusted reversibly from outside by applying external magnetic fields \cite{filipcsei2007magnetic}. This may be exploited in constructing novel damping devices \cite{sun2008study} and vibrational absorbers \cite{deng2006development}. Several theoretical studies have shown that the internal spatial particle distribution plays a qualitative role for this effect \cite{wood2011modeling,han2013field,ivaneyko2014mechanical,pessot2014structural}. 

Furthermore, applying time-dependent external magnetic fields can induce deformations, which makes the materials candidates for the use as soft actuators \cite{bohlius2004macroscopic,zimmermann2006modelling,filipcsei2007magnetic}. Related to this feature, it has been demonstrated theoretically that the spatial particle arrangement in the materials has a qualitative impact on the magnetostrictive behavior \cite{stolbov2011modelling,gong2012full,zubarev2013magnetodeformation}. 

Apart from that, quick remagnetizations of the magnetic particles by an alternating external magnetic field can lead to local heating. The effect is due to hysteretic losses in the dynamic magnetization processes. It can be used for hyperthermal cancer treatment \rev{when magnetic particles are embedded into tumor tissue} \cite{lao2004magnetic,hergt2006magnetic}. 
\rev{Ferrogels, which likewise feature magnetic particles embedded in a gel-like matrix, can serve as model systems to investigate some of the aspects that become important during this form of medical treatment.}

In all these processes, dynamic modes of the materials are excited. This happens via the time-dependence of the applied mechanical deformations and external magnetic fields. Different modes will dominate depending on the type of external stimulus. In the described situation there are two major differences when compared to the classical picture of phonon modes in conventional solids \cite{ashcroft1976solid}: we expect the dynamics of the magnetic particles to be mainly of the relaxatory kind, and the particle arrangement is not that of a regular crystalline lattice. 

A natural goal is to optimize the materials in view of their applications. For this purpose, it is important to understand if and how the dynamic modes are determined by internal structural properties and by external magnetic fields. 
\rev{So far, a macroscopic continuum theory for the dynamic response of magnetic gels has been developed using a hydrodynamic-like symmetry-based approach \cite{jarkova2003hydrodynamics,bohlius2004macroscopic}. However, particle-resolved studies that connect the dynamic material behavior to structural properties on the magnetic particle level are missing.} 
\rev{Our investigations in the following are a first step into this direction. }

\rev{In the next section, we review the simplified dipole-spring model that we recently introduced to investigate equilibrium ground states of simple model systems \cite{annunziata2013hardening}. We expand it by formulating the corresponding relaxation dynamics. Our approach contains memory terms to include a possible orientational coupling of the magnetic particles to the polymer network \cite{annunziata2013hardening}. We then demonstrate and analyze the impact of three different factors on the dynamic relaxatory modes. First, the orientational memory can qualitatively impact the appearance of the materials, which also influences the dynamic modes. This is demonstrated for the illustrative example of a short linear magnetic chain in Sec.~\ref{sec:memory}. Second, the spatial distribution of the magnetic particles is important. We depict this fact using simple symmetric lattice cells in two spatial dimensions in Sec.~\ref{sec:distribution}. Third, the mode structure can be influenced by an external magnetic field. This is highlighted for a spatial particle distribution that was extracted from the cross-section of a real experimental sample in Sec.~\ref{sec:field}. The results are summarized in Sec.~\ref{sec:conclusions}. }

\section{Dynamic dipole-spring model}\label{sec:model}


Our ambition in this paper is to qualitatively demonstrate that the relaxation dynamics can be influenced by three different factors: orientational memory, spatial distribution of the magnetic particles, and external magnetic fields. For this purpose, we employ a minimal dipole-spring approach that includes all these ingredients. 

We use the recently introduced model energy to describe the state of a ferrogel \cite{annunziata2013hardening}, 
\begin{eqnarray}
E &=& \frac{\mu_0}{4 \pi} \sum_{i,j=1, i < j}^N \frac{{\bf m}_i 
    \cdot {\bf m}_j - 3({\bf m}_i \cdot {\bf \hat{r}}_{ij})({\bf m}_j 
    \cdot {\bf \hat{r}}_{ij})}{r^3_{ij}}
    \nonumber\\
&&{} \hspace{-.8cm}{+\frac{k}{2} \sum_{\langle i,j \rangle} 
    \left( r_{ij} - r_{ij}^{(0)} \right)^2}
    {+D \sum_{\langle i,j \rangle} \left( {{\bf \hat{m}}_i 
    \cdot {\bf \hat{r}}_{ij}} - {{\bf \hat{m}}_i^{(0)} 
    \cdot {\bf \hat{r}}_{ij}^{(0)}} \right)^2} \nonumber\\
&&{} \hspace{-.8cm}{+\tau \sum_{\langle i,j \rangle} 
    \Big(  [{\mathbf{\widehat{{\bf m}_{\textnormal{$i$}} 
    \times {\bf r}_{\textnormal{$ij$}}}}}] \cdot 
    [{\mathbf{\widehat{{\bf m}_{\textnormal{$j$}} 
    \times {\bf r}_{\textnormal{$ij$}}}}}] }  \nonumber\\[-.3cm]
&&  \qquad {}-  [{\mathbf{\widehat{{\bf m}_{\textnormal{$i$}}^{\textnormal{(0)}} 
    \times {\bf r}_{\textnormal{$ij$}}^{\textnormal{(0)}}}}}] 
    \cdot [{\mathbf{\widehat{{\bf m}_{\textnormal{$j$}}^{\textnormal{(0)}} 
    \times {\bf r}_{\textnormal{$ij$}}^{\textnormal{(0)}}}}}]  \Big)^2. 
\label{eq_E}
\end{eqnarray}
Here, each of the $N$ magnetic particles carries a magnetic dipolar moment $\mathbf{m}_i$ and is located at position $\mathbf{r}_i$  ($i=1,...,N$). The distance vectors are $\mathbf{r}_{ij}=\mathbf{r}_j-\mathbf{r}_i$. For any vector $\mathbf{x}$ we use the abbreviations $x=\|\mathbf{x}\|$ and $\mathbf{\hat{x}}=\mathbf{x}/x$. All quantities with the superscript $^{(0)}$ refer to a memorized state imprinted into the material during its synthesis. We denote the sum over a limited number of close neighbors by angular brackets $\langle i,j \rangle$. 

The first line of Eq.~(\ref{eq_E}) contains the long-ranged dipolar interactions. Next, we model the elastic properties of the embedding polymer matrix by effective Hookean springs between the magnetic particles. $k$ is the spring constant. Both remaining terms include a simple form of orientational memory of the dipolar orientations: the term with the coefficient $D$ penalizes rotations of the dipole moments towards the connecting line between magnetic particles; $\tau$ penalizes relative rotations of the dipolar moments around these connecting lines, typically involving torsional deformations of the polymer matrix. See Ref.~\cite{annunziata2013hardening} for further explanations. 
In the following we only consider situations and parameter values for which a collapse due to the dipolar attractions does not occur; we thus can neglect steric repulsion between the particles. 

All magnetic particles 
are assumed to be identical. For ferrofluids \cite{rosensweig1985ferrohydrodynamics,odenbach2003ferrofluids, huke2004magnetic,ilg2006structure} this simplifying picture could capture the experimentally observed effects correctly \cite{thurm2003particle,pop2006investigation}. 
Particularly, in our case, an identical magnitude of the dipolar moments is assumed, $m_i=m$ ($i=1,...,N$). 
Then, 
five degrees of freedom remain for each particle $i$, given by 
a five-dimensional vector $\mathbf{y}_i\equiv (\mathbf{r}_i,\mathbf{\hat{m}}_i)$. Thus the relaxation dynamics of the system follows as the $5N$-dimensional coupled system of equations
\begin{equation}\label{eq_relax}
\frac{\partial \mathbf{y}_i}{\partial t} = {}-\bm{\gamma}\cdot\,\frac{\partial E}{\partial \mathbf{y}_i}, \quad i=1,...,N. 
\end{equation}
Here, our final simplifying assumption is that the relaxation rate tensor $\bm{\gamma}$ is diagonal and the same for all particles. Rescaling all lengths by an appropriate distance $l_0$, the positional relaxation rates can be adjusted to the angular ones, so that we obtain $\bm{\gamma}=\gamma\mathbf{I}$, with $\mathbf{I}$ the unity matrix. In all that follows, we measure time in units of $(\gamma k l_0^2)^{-1}$, $D$ and $\tau$ in units of $kl_0^2$, as well as the magnetic moment $m$ in units of $[kl_0^5/\mu_0]^{1/2}$. 

We linearize Eqs.~(\ref{eq_relax}) with respect to small deviations $\mathbf{\delta} \mathbf{y}_i$ from the energetic ground state. \rev{The resulting system of linearized dynamic equations is rather lengthy and listed in the Appendix. We insert an ansatz $\mathbf{\delta} \mathbf{y}_i =  \mathbf{\delta} \mathbf{y}_{0,i} e^{\lambda t}$ into the resulting system of linearized dynamic equations. Denoting by $\mathbf{\delta}\mathbf{y}$ the vector composed of all $\mathbf{\delta}\mathbf{y}_i$, the resulting system of equations can be written in the form $\lambda\mathbf{\delta}\mathbf{y}=\mathbf{M}\cdot\mathbf{\delta}\mathbf{y}$, with $\mathbf{M}$ the force matrix. Therefore, the relaxation rates $\lambda$ follow as the eigenvalues of this matrix, whereas its eigenvectors characterize the nature of the corresponding relaxatory dynamic modes. More precisely, the eigenvectors describe the spatial displacements and magnetic reorientations $\mathbf{\delta}\mathbf{y}_i=(\mathbf{\delta}\mathbf{r}_i,\mathbf{\delta}\mathbf{m}_i)$ of all particles $i=1,...,N$ during the corresponding dynamic mode. These eigenvalues and eigenvectors are obtained by standard numerical methods \cite{press1992numerical}. In our overdamped system, the relaxation rates together with the relaxatory modes characterize the dynamic behavior. }

\section{Impact of orientational memory}\label{sec:memory}


To demonstrate that the orientational memory has a qualitative impact, it is sufficient to consider a one-dimensional particle arrangement. For such a straight magnetic chain 
we had previously observed three qualitatively different energetic ground states \cite{annunziata2013hardening}. They occur for a memorized direction $\mathbf{\hat{m}}^{(0)}_i$ oblique to the chain axis and depend on the strength of the orientational memory $(D,\tau)$: 
we obtain a ``ferromagnetic'' state with all magnetic moments aligned along the chain (small $D$); an ``antiferromagnetic'' state with obliquely oriented magnetic moments rotated around the chain by $\pi$ between neighboring particles (large $D$, small $\tau$); and a ``spiral''-like arrangement with the rotation angle smaller than $\pi$ (large $D$, large $\tau$). 

\begin{figure}
  \begin{center}
    \includegraphics[width=8.6cm]{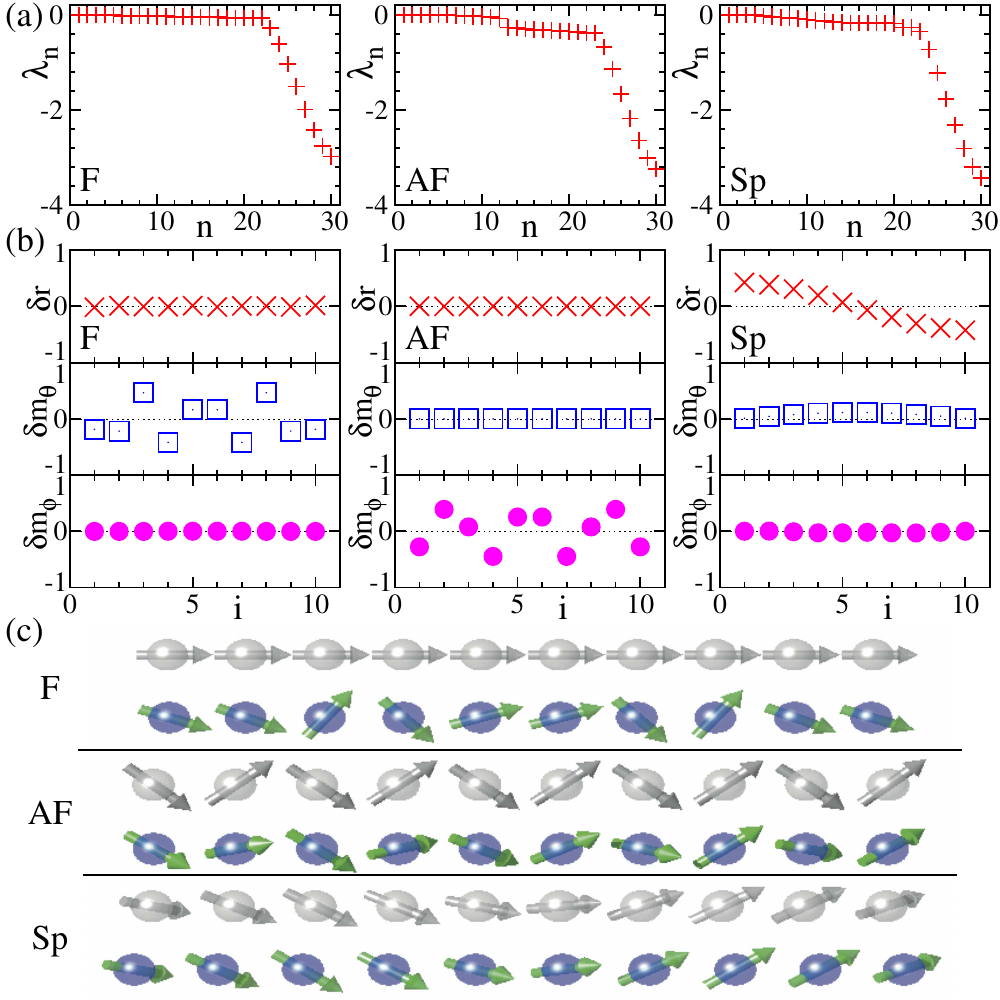}
    \caption{
    Dynamic relaxatory behavior for three different linear elastic chains of $N=10$ magnetic particles of $m=1.68$. The chains differ by orientational memory $(D,\tau)$ leading to qualitatively different energetic ground states: ferromagnetic ``F'' ($D=0.1$, $\tau=0.04$), anti-ferromagnetic ``AF'' ($D=0.6$, $\tau=0.0004$), and spiral-like ``Sp'' ($D=0.6$, $\tau=0.04$). (a) Dynamic relaxation spectra, where $n$ labels the modes. (b) Example of a characteristic eigenmode ($n=8$) that appears very differently in the three cases due to the varying orientational memory. $i$ labels the particles, $\delta r$ denotes displacements along the chain axis, $\delta m_{\theta}$ and $\delta m_{\phi}$ mark the angular deviations of the magnetic moments in spherical coordinates. (c) Illustration of the three different energetic ground states (light gray) and the resulting different modes $n=8$ as characterized in (b). In all cases the lengths of the unstrained linking springs between the particles are $r_{ij}^{(0)}=2$. 
    } 
  \label{fig_chain}
  \end{center}
\end{figure}

For illustration, we here consider a finite straight chain of only $N=10$ particles. It is characterized by an equal orientation of all memorized $\mathbf{m}_i^{(0)}$ with an angle $\sphericalangle(\mathbf{m}_{i}^{(0)},\mathbf{r}_{ij}^{(0)})=\pi/4$, the pairs $\langle i,j \rangle$ in Eq.~(\ref{eq_E}) denoting nearest neighbors. We consider three different strengths of orientational memory $(D,\tau)$ that lead to the three different ground states mentioned above, 
see further Fig.~\ref{fig_chain}. 

We determined the corresponding relaxation spectra and depict them in Fig.~\ref{fig_chain}~(a). The more negative the eigenvalue $\lambda$, the quicker the corresponding mode relaxes. We order the modes by decreasing $\lambda$. First the zero-modes of global translation along and global rotation around the chain axis are obtained. The subsequent plateau of slowly decreasing relaxation rates  
mainly contains dynamic modes dominated by rotational relaxation, see Ref.~\cite{suppl} for details. 
\rev{At the end of this plateau, there is an obvious kink in the spectral curves and the relaxation rates start to significantly decrease. For these modes, the relaxation becomes significantly quicker. Those are the modes that are dominated by longitudinal compressive and dilative displacements along the chain with higher wave numbers, again see Ref.~\cite{suppl} for details. That is, these modes can quickly decay by repositioning within small localized groups of particles implying that a collective rearrangement correlated along the whole chain is not necessary, which makes those processes faster.}
In the antiferromagnetic case, we find a specific step within the plateau region. It separates modes dominated by dipolar rotations first around and second towards the chain axis. 
As Figs.~\ref{fig_chain}~(b) and (c) show, the orientational memory can lead to qualitative differences in the nature of corresponding modes. The complete table illustrating all occurring modes is included in Ref.~\cite{suppl}. 

\rev{In the above considerations, our limitation to a relatively short chain of $N=10$ particles is due to illustrative purposes only. The differences in the spectra in Fig.~\ref{fig_chain} and in Ref.~\cite{suppl} solely result from the varying orientational memory that lead to the ferromagnetic, anti-ferromagnetic, and spiral-like ground states. Analogous results follow for significantly longer chains. Likewise, there are no qualitative differences between chains of odd and even numbers of magnetic particles for $N\geq10$ and otherwise identical parameter values.}

\rev{Summarizing, we have demonstrated the influence of the orientational memory on the dynamics for a one-dimensional spatial arrangement of the magnetic particles. Real three-dimensional bulk samples can contain such chain-like aggregates \cite{collin2003frozen,varga2003smart,filipcsei2007magnetic,borbath2012xmuct,gunther2012xray}. If the distances between the chains are large enough so that the interaction between them can be neglected \cite{zubarev2013effect}, the dynamic properties of the single chains will have a strong impact on the overall behavior. Nevertheless, the orientational memory should also become important in other cases of more isotropic particle distributions, a topic that shall be investigated further in the future. The orientational memory in our model is encoded by the parameters $D$ and $\tau$. In reality, it can for example be tuned during synthesis by the way of embedding the magnetic particles in the polymer matrix. For instance, rotations of elongated magnetic particles \cite{roeder2012shear} are hindered when compared to spheres, and magnetic particles that are actually part of the network due to chemical surface functionalization \cite{frickel2011magneto,messing2011cobalt} experience permanent restoring torques under reorientation \cite{weeber2012deformation}.
}

\section{Effect of spatial particle distribution}\label{sec:distribution}


Next, we show that the spatial distribution of the magnetic particles has an obvious impact on the relaxation dynamics. For this purpose, it is sufficient to concentrate on a two-dimensional particle arrangement. 
We consider a system without orientational memory of the dipoles, i.e. $D=0$ and $\tau=0$ in Eq.~(\ref{eq_E}). Instead, we assume that a sufficiently strong external magnetic field orients all magnetic dipoles perpendicular to the two-dimensional layer. 
Due to the above rescaling, the only remaining system parameter is the rescaled magnitude $m$ of the dipole moments. It characterizes the ratio between magnetic and elastic contributions to the system energy. 

For illustration, we consider small regular arrangements of different lattice structures and only $N=9$ particles. Of course much larger arrangements can be evaluated but not as easily be displayed. In our examples, the textures are of initially quadratic, rectangular, and hexagonal lattice structure. 
\begin{figure}
  \begin{center}
    \includegraphics[width=8.6cm]{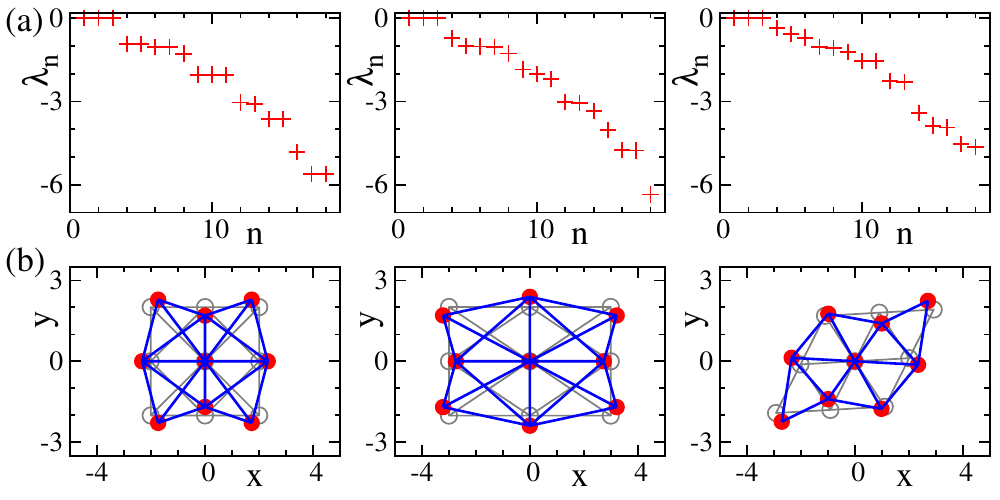}
    \caption{
    Dynamic relaxatory behavior of (from left to right) a small quadratic, rectangular (aspect ratio 2:3), and hexagonal lattice of $N=9$ particles. Magnetic moments are oriented perpendicular to the plane and of magnitude $m=1$. (a) Changes in the relaxation spectra for the three different particle distributions. (b) Different appearance of an example mode ($n=5$) for the three lattices (undeformed energetic ground states indicated in light gray). 
In all cases the lengths of the unstrained linking springs between the particles are $r_{ij}^{(0)}=2$, except for the long edges of the rectangular lattice, where they are $r_{ij}^{(0)}=3$. 
    }  
  \label{fig_lattices}
  \end{center}
\end{figure}

We display the relaxation spectra for the three different lattice structures in Fig.~\ref{fig_lattices}~(a). Since the orientations of the magnetic moments are fixed by the strong external magnetic field, all modes are solely determined by relaxations of the particle positions. In all cases, three zero modes are observed corresponding to global spatial translations and rotations. For the higher modes, the different lattice structures lead to different magnitudes of corresponding relaxation rates. Also the nature of the relaxatory modes significantly depends on the spatial particle distribution. One example is illustrated by Fig.~\ref{fig_lattices}~(b). A complete illustration of all relaxatory modes for each lattice is again included in Ref.~\cite{suppl}.

\section{Influence of an external magnetic field}\label{sec:field}


Finally, we demonstrate that an external magnetic field can change the dynamic relaxatory behavior. This is particularly important from an application point of view because it allows to tune the dynamic properties of the materials in a non-invasive way from outside. 

We consider the same set-up as above for the regular lattices. 
Now, however, there are 
${N=969}$ 
particles and their spatial distribution does not follow a regular lattice structure. In particular, to make the connection to real systems, we use a real experimental sample and extract the particle positions as an input for our study. 

\begin{figure}
  \begin{center}
    \includegraphics[width=8.6cm]{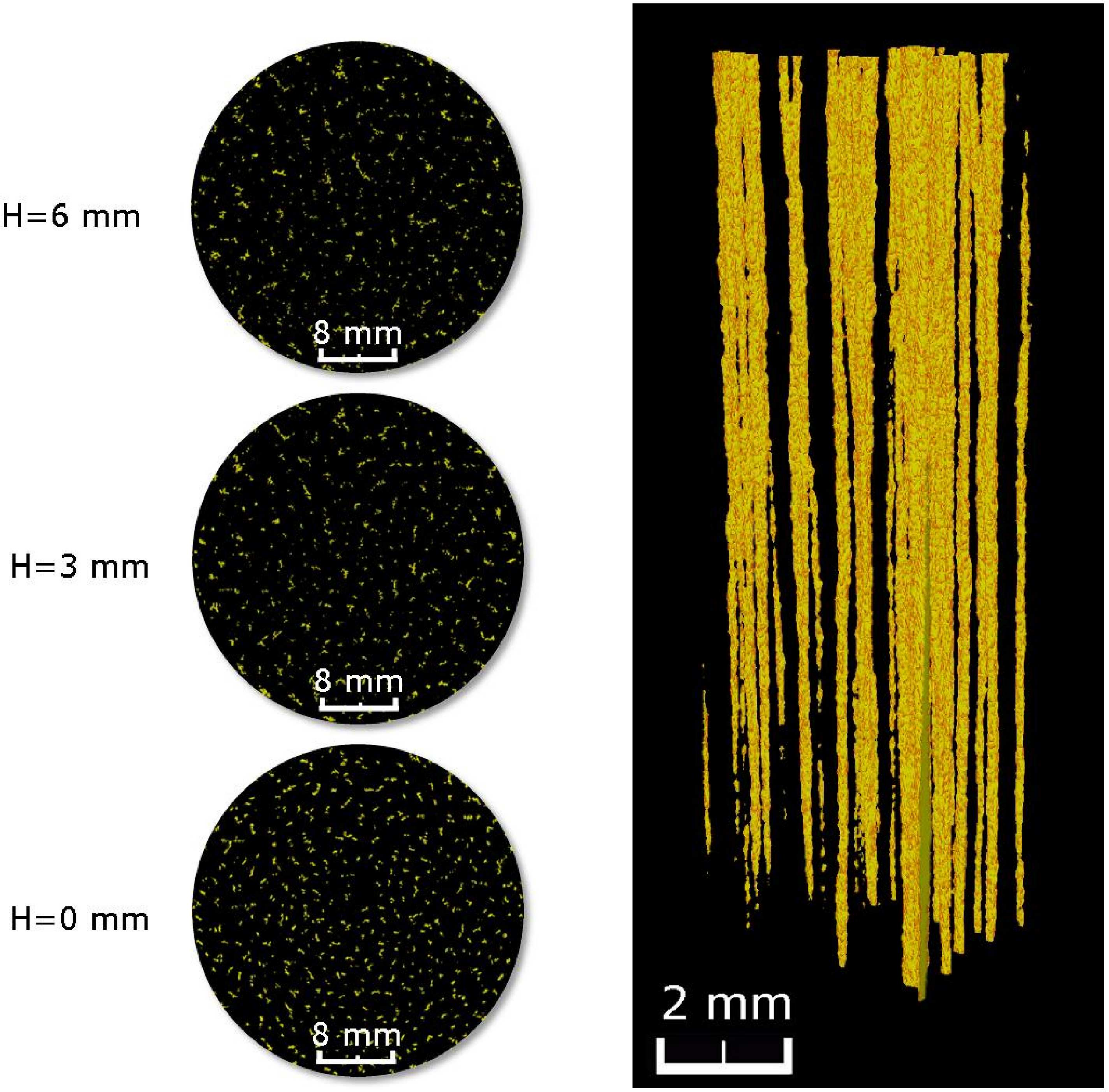}
    \caption{
      \rev{Chain-like structures observed by X-ray microtomography in the experimental sample referred to in the main text \cite{gunther2012xray}. On the left-hand side, three cross-sectional images at different heights H from the base of the sample are depicted. Bright spots label the positions of magnetic particles. On the right-hand side, a three-dimensional reconstruction of the chain-like structures formed by the magnetic particles in the sample is shown. For details of the data acquisition see Ref.~\cite{gunther2012xray}.
      Taken from Ref.~\cite{gunther2012xray}, Fig.~5. \copyright~IOP Publishing. Reproduced by permission of IOP Publishing. All rights reserved. }
    }  
  \label{fig_xray}
  \end{center}
\end{figure}

The experimental sample was extensively characterized in Ref.~\cite{gunther2012xray}. It is \rev{a two-component silicone elastomer} of cylindrical shape with a diameter of $3$~cm \rev{and a height of $1.5$~cm}. Furthermore, it contains $4.6$~wt\% of \rev{magnetically soft} iron particles, the average size of which is around $35$~$\mu$m. During the synthesis of the elastomer, a strong homogeneous external magnetic field \rev{of $220$~kA$/$m} was applied parallel to the cylinder axis. 
This resulted in the formation of linear chains of the magnetic particles spanning the whole sample parallel to the cylinder axis. The chains were resolved by X-ray microtomography \cite{gunther2012xray}, \rev{the result of which is displayed in Fig.~\ref{fig_xray}}. Cross-sectional images in planes perpendicular to the cylinder axis are available, \rev{see the left column of Fig.~\ref{fig_xray},} and contain information about the chain positions \cite{gunther2012xray}. 

%
\rev{To first approximation, due to the linear chain-like aggregates that are all oriented in the same direction, the structure at intermediate height of the sample is translationally invariant along the cylinder axis. The exact positions and sizes of individual magnetic particles in the sample could not be resolved. We consider by our model the situation within one cross-sectional layer cut out from the sample at intermediate height H. In our example, we choose the cross-section at height H=3~mm in Fig.~\ref{fig_xray}. }

\begin{figure*}
  \begin{center}
    \includegraphics[width=17.cm]{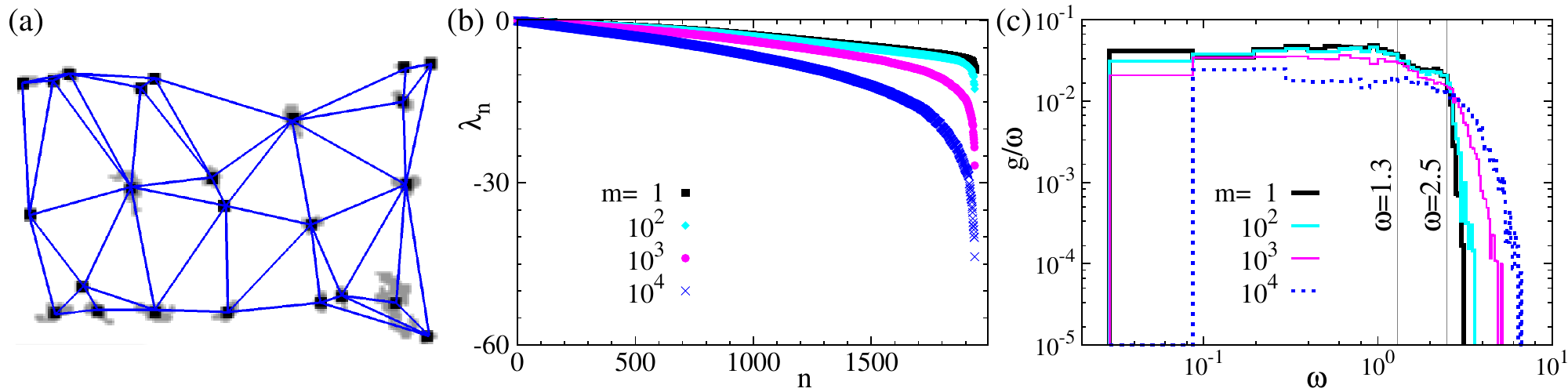}
    \caption{
      Tunability of the dynamic behavior by an external magnetic field oriented perpendicular to the plane and affecting the magnetic moments. (a) Positions of the magnetic moments are extracted from the X-ray microtomographic cross-sectional image of an anisotropic real experimental sample \cite{gunther2012xray} displayed for H=3~mm on the left-hand side of Fig.~\ref{fig_xray}. Only a fraction of the image is shown for illustration. Gray areas correspond to the microtomographic spots. (b) Tunability of the spectrum by changing the magnetization.  
      (c) The density of dynamic modes gets shifted in the frequency direction by adjusting the magnetization. \rev{Dynamic modes for $\omega\approx1.3$ and $\omega\approx2.5$ ($m=1$) are illustrated in Fig.~\ref{fig_expmodes}~(b) and (c), respectively.} [The tomography data in panel (a) are taken from Ref.~\cite{gunther2012xray}, Fig.~5 (H=3~mm), \copyright~IOP Publishing. Reproduced by permission of IOP Publishing. All rights reserved.]
    } 
  \label{fig_exp}
  \end{center}
\end{figure*}

Each spot in the cross-sectional tomography data identifies magnetic chain particles. We extracted by image analysis the centers of these spots, see Fig.~\ref{fig_exp}~(a). 
Then, in our model, we place one particle on each center, carrying a magnetic moment $\mathbf{m}$ oriented perpendicular to the plane. 
Finally, as shown in  Fig.~\ref{fig_exp}~(a), the area between the particles is tessellated by Delaunay triangulation. 
We insert elastic springs along the edges of the resulting triangles, which sets the pairs described by $\langle i,j \rangle$ in Eq.~(\ref{eq_E}). 
\rev{Magnetic interactions are still considered between all pairs of magnetic particles in the plane.} 
In this way, we model the physics of one cross-sectional layer of the real system. 
Since the magnetic particles in the experimental sample are not covalently bound to the polymer matrix \cite{frickel2011magneto,messing2011cobalt}, and since 
the magnetic moments are 
perpendicular to the plane, the orientational memory terms in Eq.~(\ref{eq_E}) do not play a role. 

For large enough particle sizes, the magnetization of the particles and thus the magnitude of their magnetic moments can be tuned by the strength of an external magnetic field. 
\rev{We consider this external magnetic field perpendicular to the plane, i.e.\ along the linear chains formed by the magnetic particles in the real sample. This has two reasons. First, we know from the procedure of synthesis that such a magnetic field orients the magnetic moments perpendicular to the plane and maintains the axial symmetry of the sample. And second, in the static case, the largest degree of tunability of the compressive elastic modulus was achieved when the magnetic field was oriented parallel to the anisotropy direction \cite{filipcsei2007magnetic}. A similar dependence may also hold in the dynamic case.} 
To keep the description general and simple, we do not consider specific magnetization laws but study the relaxation dynamics directly as a function of the magnitude of the resulting dipolar magnetic moment $m$. 

As is obvious from Fig.~\ref{fig_exp}~(b), the dynamic relaxation spectra can be tuned by adjusting 
$m$. We checked that the chosen values correspond to external magnetic field strengths that can be realized experimentally. In our geometry, the magnetic interactions within the plane are purely repulsive. 
\rev{Moreover, as can also be seen from Figs.~\ref{fig_xray} and \ref{fig_exp}~(a), the sizes of the spots detected by X-ray microtomography in the cross-sectional layers is not homogeneous. In a variant of our approach, we varied the strengths of the magnetic moments proportionally to the area of the detected spots. However, this did not qualitatively influence our results.}

\begin{figure*}
  \begin{center}
    \includegraphics[width=17.cm]{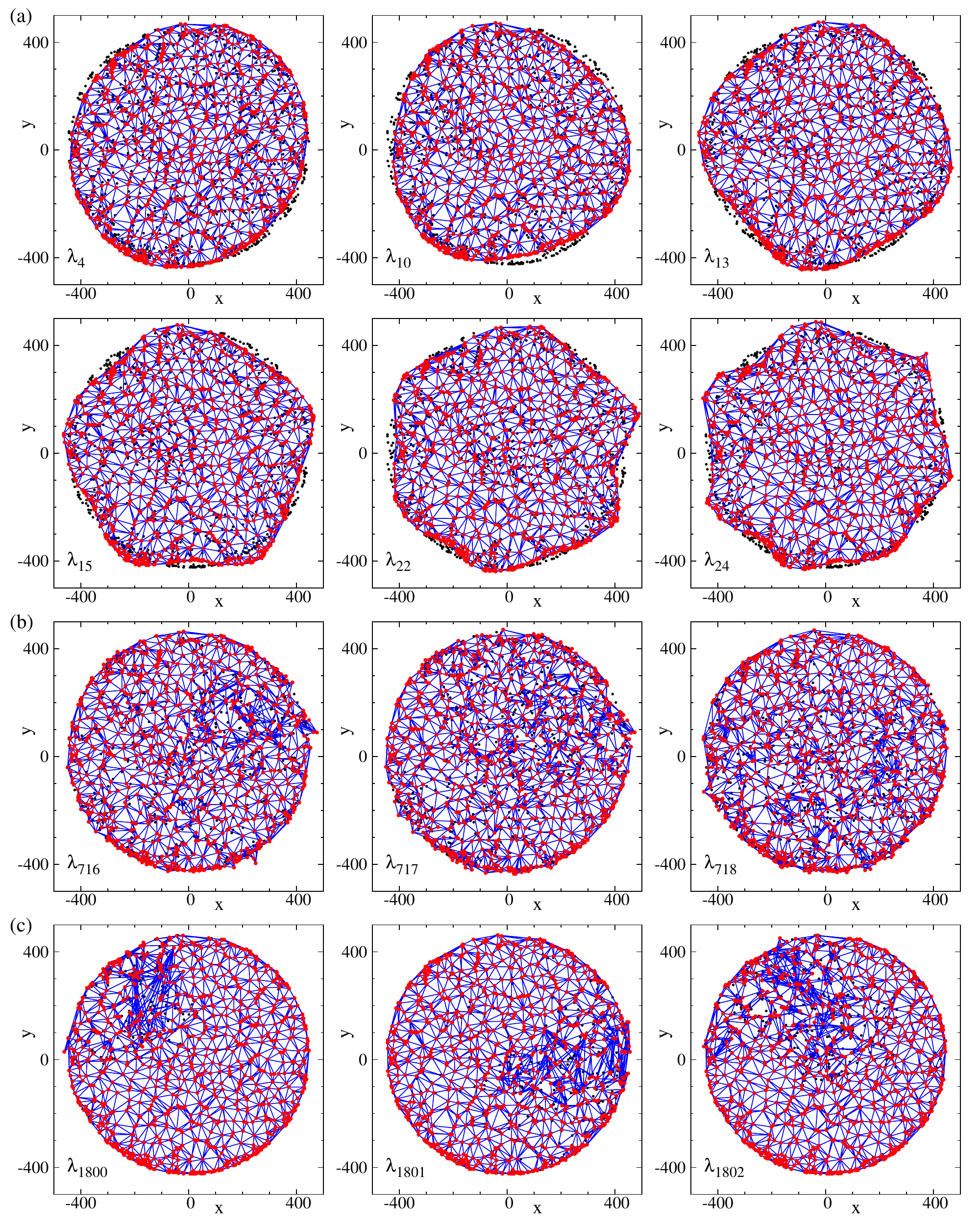}
    \caption{
      \rev{Illustration of dynamic relaxational modes for the $969$-particle planar irregular lattice extracted from the experimental sample. Colored illustrations of the deformed lattices for each mode are superposed to the black undistorted lattice corresponding to the energetic ground state.
    (a) Examples of lower modes show the expected global collective deformations, here of elliptic ($\lambda_{4}$), triangular ($\lambda_{10}$), quadratic ($\lambda_{13}$), pentagonal ($\lambda_{15}$), hexagonal ($\lambda_{22}$), and heptagonal ($\lambda_{24}$) shape. Selected eigenmodes (b) around the end of the ``Debye plateau'' ($\omega \approx 1.3)$ and (c) around the small hump that might be connected to a ``boson peak'' ($\omega \approx 2.5$), cf.\ Fig.~\ref{fig_exp}~(c), show a much more localized character. 
The initial spring lengths $r^{(0)}_{ij}$ were set according to the values extracted from the experimental sample, while the magnetic moment was chosen as $m=1$. }
    } 
  \label{fig_expmodes}
  \end{center}
\end{figure*}

\rev{Fig.~\ref{fig_expmodes} displays several illustrative example modes from the spectrum for $m=1$ in Fig.~\ref{fig_exp}~(b). Black dots mark the initial positions of the magnetic particles, whereas the overlayed lattice shows the deformed state. The directions and relative magnitudes of the displacements of the individual particles are obtained from the eigenvectors calculated as described at the end of Sec.~\ref{sec:model}.} 

There are two major differences when compared to the classical phonon modes in crystalline solids \cite{ashcroft1976solid}. First, our dynamics is overdamped \cite{ivlev2012complex}. 
\rev{Therefore, we here focus on the relaxational spectra determined from the corresponding relaxation rates $\lambda_n$.}
Second, our lattice is irregular. 
Nevertheless, the situation is typically discussed in terms of the mode density $g(\omega)$ in frequency space following the notation of the classical phonon picture of non-overdamped oscillations \cite{ashcroft1976solid}. 
\rev{The frequencies $\omega_n$ of these oscillations in the classical phonon picture would be determined from the same force matrix as the one that we find from the right-hand side of Eq.~(\ref{eq_relax}). However, on the left-hand side of Eq.~(\ref{eq_relax}), the phonon oscillations would imply a second time derivative. The two quantities that appear on the left-hand side in these two different cases are related by $\omega_n \sim \sqrt{\left| \lambda_n \right|}$. Since it is common to plot the mode density $g(\omega)$ in frequency space, we adhere to this convention.}

At not too high frequencies that correspond to long-scale collective dynamics, the plane-wave picture should still apply. 
In fact, in this regime, a behavior of $g(\omega)$ in accordance with the classical Debye picture \cite{ashcroft1976solid} was obtained for disordered structures \cite{kaya2010normal}. 
\rev{Likewise, we observe here for our two-dimensional disordered solid a ``Debye plateau'' of the function $g(\omega)/\omega$ 
in Fig.~\ref{fig_exp}~(c) at not too high frequencies. 
Example modes at the low-frequency end of the spectrum indeed are related to long-scale collective dynamics, as demonstrated in Fig.~\ref{fig_expmodes}~(a).}

However, instead of a pure drop of $g(\omega)$ at higher frequencies, a typical ``boson peak'' 
can develop in disordered systems \cite{kaya2010normal}, the origin of which is still under debate \cite{shintani2008universal}. 
In our example of a two-dimensional disordered solid, the curve for $g(\omega)$ in Fig.~\ref{fig_exp}~(c), before it drops at the end of the plateau, shows a small hump. It is not possible to decide on the basis of our limited \rev{statistical} data whether this is the signature of a ``boson peak'' \rev{in our non-glassy system}. What does become obvious \rev{from Figs.~\ref{fig_expmodes}~(b) and (c) is that the higher-frequency modes are significantly more localized. This explains their higher relaxation rates: large-scale collective motion is not necessary to relax them.}

Most significant for our present purposes is the observation in Fig.~\ref{fig_exp}~(c) that the spectral density $g(\omega)$ can be shifted in frequency direction by adjusting $m$ through an external magnetic field. This is an important ingredient from the application point of view. It allows to adjust the relaxation time reversibly in response to varying environmental conditions. 
\rev{We recall that the fraction of the magnetic component in our experimental sample was only $4.6$~wt\%. Significantly higher contents of magnetic particles can be realized. It turns out that already after halving all distances in our sample plane, switching $m$ from $10$ to $10^3$ is sufficient to achieve a similar shift in the spectrum as the one occuring in Fig.~\ref{fig_exp}~(c) between $m=1$ and $m=10^4$. This underlines the potential of the magnetic interaction as an effective control parameter for the dynamic behavior.} 
In combination with the established phononic properties of colloidal systems \cite{penciu2003phonons,baumgartl2008phonon,still2011collective}, this mechanism could provide a route to tunable sound absorbers.

\section{Conclusions}\label{sec:conclusions}


Summarizing, we have demonstrated that the dynamic behavior of ferrogels and magnetic elastomers can be tailored and adjusted by at least three factors: first, by the magneto-elastic coupling and orientational memory; 
second, by the particle distribution; 
and third, during application, by external magnetic fields. 
Thus we can forecast how microscopic details, e.g.\ the orientational coupling of the magnetic particles to their polymeric environment, affect aspects of the dynamic material properties. 
\rev{There are of course several further factors that determine our model parameters and in this way influence the relaxation behavior. For example, these could be the content of magnetic particles, the temperature during application, the degree of crosslinking, or the degree of swelling of the materials. The impact of these parameters should be analyzed both experimentally and theoretically in the future. On the experimental side, aspects of the dynamic relaxation properties can be inferred, for instance, from dielectric relaxation studies \cite{frickel2011magnetic} or nanorheology \cite{roeben2014magnetic}. Having all these tuning parameters at hand, it should be possible to adjust the dynamic properties to the requested applicational need.} 

\rev{To our knowledge, investigating aspects of the dynamic material behavior on the level of the magnetic particle distributions represents a new direction in the field. We hope that our study can stimulate further, more detailed, and more quantitative theoretical and simulation work in this context. Naturally, the extension to three spatial dimensions is an important next step. Our main goal here was to outline for simple one- and two-dimensional model cases the different factors that can influence the dynamics of the systems. To allow for quantitative predictions on the dynamic behavior of real samples, three-dimensional analyses will be mandatory in most situations. On the experimental side, for example the differences between isotropic and uniaxial ferrogels should be analyzed concerning dynamic properties. All of these questions are of high practical relevance in view of the dynamic applications. For instance, response and relaxation times determine the range of usability of ferrogels as the basis of the above-mentioned novel damping devices \cite{sun2008study}, vibrational absorbers \cite{deng2006development}, or soft actuators \cite{zimmermann2006modelling,filipcsei2007magnetic}. }

Our analysis represents a first step towards an optimization of the dynamic behavior of magnetic gels. 
Theory and simulations could assist this process by identifying particle properties and structural arrangements that lead to the requested characteristics. 
\rev{A further investigation to connect our approach to directly experimentally measured quantities such as the dynamical susceptibilities is currently underway \cite{pessot2014unpublished}.} 
We hope that our study can stimulate further investigations to support the design of these fascinating materials and optimize their tunable dynamic properties.

\begin{acknowledgments}
The authors thank the Deutsche Forschungsgemeinschaft for support of this work through the priority program SPP 1681. 
M.T.\ was supported by the JSPS Core-to-Core Program ``Non-equilibrium dynamics of soft matter and information'' and a JSPS Research Fellowship.
\end{acknowledgments}

\appendix*

\section{Dynamic equations for the relaxational behavior}

Here we list the complete expressions for the linearized equations characterizing the relaxation dynamics and following from Eq.~(\ref{eq_relax}). 
In this way, the relaxation dynamics of small deviations $\mathbf{\delta} \mathbf{y}_i$ from the energetic ground state is obtained, where $\mathbf{y}_i\equiv (\mathbf{r}_i,\mathbf{\hat{m}}_i)$ and $i=1,...,N$ labels the particles: 
\begin{equation}
\frac{\partial \delta\mathbf{y}_i}{\partial t} = {}-\gamma \sum_{j=1}^N \mathcal{L}_{ij} \delta\mathbf{y}_j.
 \label{eq.suppl:1}
\end{equation}
For simplicity, we only show the formulae for the one-dimensional chain and for the two-dimensional planar particle arrangements considered in the main text.
In the latter case, we assume that the orientation of the magnetic dipoles is fixed perpendicular to the plane. This can, for example, be achieved by a strong external magnetic field. 

\subsection{Linear chain-like particle arrangement}

First, for the one-dimensional chain-like aggregates, the vector $\mathbf{y}_i$ reduces to a three-dimensional vector  $\mathbf{y}_i\equiv( r_i, \theta_i, \phi_i )$. 
In our choice of coordinates, $r_i$ marks the position of the $i$th particle along the chain, whereas the two angles $\theta_i$ and $\phi_i$ represent the azimuthal and polar angles of the dipolar orientation of the particle with respect to the chain direction. 
The linearized operator in the above Eq.~(\ref{eq.suppl:1}) is separated into four parts resulting from the four contributions to the energy $E$ in Eq.~(1) of the main text: 
\begin{equation}
\mathcal{L}_{ij} = \mathcal{L}_{ij}^{\rm dip} +\mathcal{L}_{ij}^{\rm el} +\mathcal{L}_{ij}^D +\mathcal{L}_{ij}^{\tau}. 
 \label{eq.suppl:2}
\end{equation}

We start by calculating the contribution from the dipole-dipole interaction energy. 
Its diagonal components are given by
\begin{eqnarray}
&\mathcal{L}^{\rm dip}_{ii}& 
 = \frac{3 \mu_0}{4 \pi} \sum_{k \neq i} \left| r_{ik} \right|^{-3} r_{ik}^{-1} m^2 \notag\\
 &&{} \hspace{-.8cm} \times \big[ 4 r_{ik}^{-1} \big\{ \sin\theta_i \sin\theta_k \cos \left( \phi_k -\phi_i \right) -2 \cos\theta_i \cos \theta_k \big\} \hat{r} \hat{r} \notag\\
 &&{} \hspace{-.8cm} + \big\{ \cos\theta_i \sin\theta_k \cos\left( \phi_k -\phi_i \right) +2 \sin\theta_i \cos\theta_k \big\} \hat{r} \hat{\theta}_i \notag\\
 &&{} \hspace{-.8cm} + \sin\theta_i \sin\theta_k \sin \left(\phi_k -\phi_i \right) \hat{r} \hat{\phi}_i \big] \notag\\
 &&{} \hspace{-.8cm} + \frac{\mu_0}{4 \pi} \sum_{k \neq i} \left| r_{ik} \right|^{-3} m \notag\\
 &&{} \hspace{-.8cm} \times \big[ 3 r_{ik}^{-1} \big\{ \cos\theta_i \sin\theta_k \cos\left( \phi_k -\phi_i \right) +2 \sin\theta_i \cos\theta_k \big\} \hat{\theta}_i \hat{r} \notag\\
 &&{} \hspace{-.8cm} - \big\{ \sin\theta_i \sin\theta_k \cos\left( \phi_k -\phi_i \right) -2 \cos\theta_i \cos\theta_k \big\} \hat{\theta}_i \hat{\theta}_i \notag\\
 &&{} \hspace{-.8cm} + 3 r_{ik}^{-1} \sin\theta_k \sin\left( \phi_k -\phi_i \right) \hat{\phi}_i \hat{r} \notag\\
 &&{} \hspace{-.8cm} - \sin\theta_i \big\{ \sin\theta_i \sin\phi_k \cos\left( \phi_k -\phi_i \right) -2 \cos\theta_i \cos\theta_k \big\} \hat{\phi}\hat{\phi} \big] \notag\\
 \label{eq.supple:4}
\end{eqnarray}
and its off-diagonal components for $j \neq i$ by
\begin{eqnarray}
&\mathcal{L}^{\rm dip}_{i j\neq i}& = \frac{3 \mu_0}{4 \pi} \left| r_{ij} \right|^{-3} r_{ij}^{-1} m^2 \notag\\
 &&{} \hspace{-1.2cm} \times \big[ -4 r_{ij}^{-1} \big\{ \sin\theta_i \sin\theta_j \cos \left( \phi_j -\phi_i \right) -2 \cos\theta_i \cos \theta_j \big\} \hat{r} \hat{r} \notag\\
 &&{} \hspace{-1cm} + \big\{ \sin\theta_i \cos\theta_j \cos\left( \phi_j -\phi_i \right) +2 \cos\theta_i \sin\theta_j \big\} \hat{r} \hat{\theta}_j \notag\\
 &&{} \hspace{-1cm} - \sin\theta_i \sin\theta_j \sin \left(\phi_j -\phi_i \right) \hat{r} \hat{\phi}_j \big] \notag\\
 &&{} \hspace{-1cm} + \frac{\mu_0}{4 \pi} \left| r_{ij} \right|^{-3} m \notag\\
 &&{} \hspace{-1.2cm} \times \big[ -3 r_{ij}^{-1} \big\{ \cos\theta_i \sin\theta_j \cos\left( \phi_j -\phi_i \right) +2 \sin\theta_i \cos\theta_j \big\} \hat{\theta}_i \hat{r} \notag\\
 &&{} \hspace{-1cm} + \big\{ \cos\theta_i \cos\theta_j \cos\left( \phi_j -\phi_i \right) -2 \sin\theta_i \sin\theta_j \big\} \hat{\theta}_i \hat{\theta}_j \notag\\
 &&{} \hspace{-1cm} - \cos \theta_i \sin\theta_j \sin \left( \phi_j -\phi_i \right) \hat{\theta}_i \hat{\phi}_j \notag\\
 &&{} \hspace{-1cm} 
- 3 r_{ij}^{-1} \sin\theta_j \sin\left( \phi_j -\phi_i \right) \hat{\phi}_i \hat{r} + \cos\theta_j \sin\left( \phi_j -\phi_i \right) \hat{\phi}_i \hat{\theta}_j  \notag\\
 &&{} \hspace{-1cm} 
+ \sin\theta_i \cos\left( \phi_j -\phi_i \right) \hat{\phi}_i \hat{\phi}_j \big]. 
 \label{eq.supple:5}
\end{eqnarray}
Here $r_{ij} = r_j -r_i$ and $\hat{r}$ denotes the unit vector in $r$-direction, i.e.\ along the chain axis. Likewise, $\hat{\theta}_i$ and $\hat{\phi}_i$ represent the unit vectors in the $\theta$- and $\phi$-direction for the current orientation of the dipolar moment of the $i$th particle. 

After straightforward calculation, the components of the operator containing the elastic part are obtained as
\begin{equation}
\mathcal{L}^{\rm el}_{ij} = 
 \left\{
 \begin{array}{cl}
 k \sum_{\ell \in \delta\Omega_i} \hat{r} \hat{r} & \textrm{~if~} i=j, \\[.1cm]
 -k \hat{r} \hat{r} & \textrm{~if~} j \in \delta\Omega_i, \\[.1cm]
 0 & \textrm{~otherwise},
 \end{array}
 \right.
 \label{eq.suppl:3}
\end{equation}
where $\delta\Omega_i$ denotes the set of the (one or two) nearest neighbors of the $i$th particle. 

In the same way, the first contribution from the orientational memory becomes
\begin{eqnarray}
\mathcal{L}^{D}_{i i} 
 &=& -2 D m^{-1} \notag\\
 &\times& \Big[ \big\{ -\sin^2 \theta_i +\left( \cos \theta_i -\cos\theta^{(0)}_i \right) \cos\theta_i \big\} \hat{\theta}_i \hat{\theta}_i \notag\\
 && + \left( \cos\theta_i -\cos\theta^{(0)}_i \right) \sin\theta_i \cos\theta_i \hat{\phi}_i \hat{\phi}_i \Big]
 \label{eq.suppl:6}
\end{eqnarray}
and \begin{equation}
\mathcal{L}^{D}_{i j \neq i} = 0.
 \label{eq.suppl:7}
\end{equation}

Finally, the diagonal components of the linearized operator resulting from the second part of the orientational memory are calculated as 
\begin{eqnarray}
\mathcal{L}^{\tau}_{i i} 
 &=& 2 \tau \sum_{k \in \delta\Omega_i} m^{-1} \left( \sin \theta_i \right)^{-1} \notag\\
 &\times& \Big[ -\cos\theta_i \sin \left( \phi_k -\phi_i \right) \notag\\
 && \times \Big\{ \cos \left( \phi_k -\phi_i \right) -\cos \left( \phi^{(0)}_k -\phi^{(0)}_i \right)  \Big\} \hat{\theta}_i \hat{\phi}_i \notag\\
 &-& \left( \sin\theta_i \right)^{-1} \cos\theta_i \sin\left( \phi_k -\phi_i \right) \notag\\
 && \times \Big\{ \cos\left( \phi_k -\phi_i \right) -\cos\left( \phi^{(0)}_k -\phi^{(0)}_i \right) \Big\} \hat{\phi}_i \hat{\theta}_i \notag\\
 &-& \Big\{ -\cos\left( \phi_k-\phi_i \right) \cos\left( \phi^{(0)}_k -\phi^{(0)}_i \right) \notag\\
 && +\cos^2 \left( \phi_k -\phi_i \right) -\sin^2 \left( \phi_k -\phi_i \right) \Big\} \hat{\phi}_i \hat{\phi}_i \Big]. \notag\\&&
 \label{eq.suppl:8}
\end{eqnarray}
The corresponding off-diagonal components are given by
\begin{eqnarray}
\mathcal{L}^{\tau}_{ij \in \delta\Omega_i}
 &=& 2 \tau m^{-1} \left( \sin \theta_i \right)^{-1} \notag\\
 &\times& \Big\{ -\cos \left( \phi_j -\phi_i \right) \cos \left( \phi^{(0)}_j -\phi^{(0)}_i \right) \notag\\
 &+& \cos^2 \left( \phi_j -\phi_i \right) -\sin^2 \left( \phi_j -\phi_i \right) \Big\} \hat{\phi}_i \hat{\phi}_j   \notag\\&&
 \label{eq.suppl:9}
\end{eqnarray}
for pairs of nearest neighbors. Otherwise the off-diagonal components are zero, 
\begin{equation}
\mathcal{L}^{\tau}_{ij \notin \Omega_i} = 0,
 \label{eq.suppl:10}
\end{equation}
with $\Omega_i = \delta\Omega_i + \{ i \}$ in this notation.

\subsection{Planar particle arrangement}

Second, in the case of the two-dimensional plane, we assume that all dipole moments are aligned perpendicular to the plane. 
Then, since the degrees of freedom for the dipolar orientations drop out, the vector $\mathbf{y}_i$  reduces to two dimensions, i.e.\ $\mathbf{y}_i\equiv\left( x_i, y_i \right)$. Furthermore, the two terms of orientational memory characterized by the coefficients $D$ and $\tau$, vanish. 
As a result, the linearized operator in Eq.~(\ref{eq.suppl:1}) above contains only two contributions resulting from the dipolar and from the elastic part of the energy $E$ in Eq.~(1) of the main text:
\begin{equation}
\mathcal{L}_{ij} = \mathcal{L}^{\rm dip}_{ij} + \mathcal{L}^{\rm el}_{ij}.
 \label{eq.suppl:16}
\end{equation}

The operator characterizing the dipole-dipole interactions is linearized to 
\begin{eqnarray}
&\mathcal{L}^{\rm dip}_{ii}&
 = \frac{3 \mu_0}{4 \pi} \sum_{k \in\delta\Omega_i}  r_{ik}^{-7} m^2 \notag\\
 &&{} \hspace{-.8cm} \times \Big[ \big\{ 5 \left( x_k -x_i \right)^2 -r_{ik}^2 \big\} \hat{x} \hat{x} +\big\{ 5 \left( y_k -y_i \right)^2 -r_{ik}^2 \big\} \hat{y} \hat{y} \notag\\
 &&{} + 5 \left( x_k -x_i \right) \left( y_k -y_i \right) \left( \hat{x} \hat{y} +\hat{y} \hat{x} \right) \Big]
 \label{eq.suppl:14}
\end{eqnarray}
for the diagonal components and to 
\begin{eqnarray}
&\mathcal{L}^{\rm dip}_{i j\neq i}& 
 = -\frac{3 \mu_0}{4 \pi} r_{ij}^{-7} m^2 \notag\\
 &&{} \hspace{-.8cm} \times \Big[ \big\{ 5 \left( x_j -x_i \right)^2 -r_{ij}^2 \big\} \hat{x} \hat{x} +\big\{ 5 \left( y_j -y_i \right)^2 -r_{ij}^2 \big\} \hat{y} \hat{y} \notag\\
 &&{} +5 \left( x_j -x_i \right) \left( y_j -y_i \right) \left( \hat{x} \hat{y} +\hat{y} \hat{x} \right)  \Big]
 \label{eq.suppl:15}
\end{eqnarray}
for the off-diagonal components. 

For the linearized operator resulting from the elastic contribution, the diagonal components read
\begin{eqnarray}
\mathcal{L}^{\rm el}_{i i}
 &=& -k \sum_{\ell \in \delta\Omega_i} r_{i \ell}^{-1} \Big[ - \big\{ L r_{i \ell}^{-2} \left( x_{\ell} -x_i \right)^2 +r_{i\ell} -L \big\} \hat{x} \hat{x} \notag\\
 &-& L r_{i \ell}^{-2} \left( x_{\ell} -x_i \right) \left( y_{\ell} -y_i \right) \left( \hat{x} \hat{y} + \hat{y} \hat{x} \right) \notag\\
 &-& \big\{ L r_{i \ell}^{-2} \left( y_{\ell} -y_i \right)^2 + r_{i\ell} -L \big\} \hat{y} \hat{y}
 \Big].
 \label{eq.suppl:11}
\end{eqnarray}
Its off-diagonal components are obtained as
\begin{eqnarray}
\mathcal{L}^{\rm el}_{i j \in\delta\Omega_i}
 &=& -k  r_{ij}^{-1} \Big[  \big\{ L r_{ij}^{-2} \left( x_j -x_i \right)^2 +r_{ij} -L \big\} \hat{x} \hat{x} \notag\\
 &+& L r_{ij}^{-2} \left( x_j -x_i \right) \left( y_j -y_i \right) \left( \hat{x} \hat{y} + \hat{y} \hat{x} \right) \notag\\
 &+& \big\{ L r_{ij}^{-2} \left( y_j -y_i \right)^2 + r_{ij} -L \big\} \hat{y} \hat{y}
 \Big]
 \label{eq.suppl:12}
\end{eqnarray}
for nearest neighbors and otherwise as
\begin{equation}
\mathcal{L}^{\rm el}_{i j \notin\Omega_i} = 0, 
 \label{eq.suppl:13}
\end{equation}
where again $\Omega_i = \delta\Omega_i + \{ i \}$.


\end{document}